\documentstyle[seceq,preprint,epsf]{jpsj}

\def\runtitle{Effects of Umklapp Scattering on Electronic States 
in One Dimension}
\def\runauthor{Masakazu {\sc Murakami
} and Hidetoshi {\sc Fukuyama}}

\title
{
Effects of Umklapp Scattering on Electronic States 
\\in One Dimension
}

\author
{ 
Masakazu {\sc Murakami}
\footnote{e-mail address : murakami@watson.phys.s.u-tokyo.ac.jp}
and Hidetoshi {\sc Fukuyama}
}

\inst
{
Department of Physics, Tokyo University, Tokyo 113
}

\recdate
{
March 28, 1997}

\abst
{
The effects of Umklapp scattering on electronic states are studied
in one spatial dimension at absolute zero.
The model is basically the Hubbard model, where parameters 
characterizing the
normal ($U$) and Umklapp ($V$) scattering are treated independently.
The density of states is calculated in the t-matrix approximation by 
taking only the forward and Umklapp scattering into account.
It is found that the Umklapp scattering causes 
the global splitting of the density of states. 
In the presence of sufficiently
strong Umklapp scattering, a pole in the t-matrix appears 
in the upper half plane, signalling an instability towards the '$G/2-$pairing'
ordered state ($G$ is the reciprocal lattice vector), 
whose consequences are studied 
in the mean field approximation. It turns out that
this ordered state coexists with 
spin-density-wave state and also brings about Cooper-pairs. 
A phase diagram is determined in the plane of
$V$ and electron filling $n$. }

\kword
{
modified t-matrix approximation, Umklapp scattering, density of states, 
mean field approximation, 
$G/2-$pairing, $G/2-$SDW.
}

\begin{document}
\sloppy
\maketitle
\section{Introduction}
The problems of the metal-insulator transition due to the short-range 
Coulomb repulsion in the lattice system
(known as the Mott transition\cite{Mott}) 
have attracted much interest.
The Mott transition is often examined based on the Hubbard model\cite{Hubbard}
with the hopping integral $t$ 
and the on-site Coulomb repulsion $U$.
At least qualitatively, this phenomenon can be understood:  
If $U/t$ is large, electrons are kept apart each other.
Especially for a half-filled band (one electron per one lattice site),
they get completely localized because it requires large energy 
comparable with $U$ 
for them to move to another lattice site.
As a result, the charge excitation
spectrum has a gap known as the Mott-Hubbard gap and 
the system becomes an insulator.  
On the other hand if $U/t$ is small, it is energetically favorable 
for electrons to spread into the energy band and the system behaves as a metal.
Hence it is understood that  both the Coulomb repulsion and the existence
of discrete lattice 
are the keys to the Mott transition.
It was Hubbard who first succeeded in describing the qualitative feature of
such behaviors\cite{Hubbard}. 
By using the Hubbard model, he showed that the Mott-Hubbard gap 
begins to appear as $U/t$ gets larger in the
half-filled case.

If we consider the on-site Coulomb interaction $U$ term 
in the wave vector representation,
we have not only normal processes (which conserve the total momentum
of two scattering electrons) but also Umklapp processes.
Since the latter processes originate from the periodicity of the lattice 
and stand for 
the large-momentum-transfer processes, they play an important role
in the Mott transition. 

In the space dimension $d=1$, the exact solutions of the Hubbard model
have been obtained 
by using Bethe Ansatz method\cite{betheansatz1}. 
It was proved that the charge excitation 
spectrum has a gap and the system becomes an insulator 
for a half-filled band for any strength of Coulomb interaction.
In this method, however, it is not easy to obtain the physically clear-cut 
understanding of the
origin of the Mott-Hubbard gap. On the other hand,
it was shown by the bosonization scheme\cite{Emery1}
that the Umklapp scattering produces this gap in a half-filled 
band\cite{Emery2,Giamarchi}. Moreover,
characteristic features of the charge excitations
in one-dimensional doped Mott insulators, especially the crossover between
the metallic Drude type of excitation and the Mott-Hubbard gap, have been
clarified\cite{Morikun} 
by mapping the model first to 
the g-ology model\cite{Emery1,Solyom,1dconductor}, then to the phase 
Hamiltonian\cite{1dconductor} and finally to the massive
Thirring model.
It has also been disclosed that the Mott-Hubbard gap and other associated
anomalies (the divergence of the compressibility near half filling, etc.)
originate from Umklapp processes
and how the system begins to show anomalies
as the system approaches half filling
\cite{Giamarchi,Morikun,Emery3,Schulz}.

The emergence of the Mott insulator has also been clarified
in the case of $d=\infty$ and it was also shown that
the density of states has a gap together with the existence of quasi-particle
peak before the onset of Mott transition\cite{dinf}.
In $d=\infty$, however, we cannot distinguish Umklapp processes 
from normal ones because of the irrelevance of the momentum conservation. 

From this point of view, it will be important to clarify the roles of 
the Umklapp scattering especially near half filling. 
In this paper, we examine the effects of Umklapp processes 
in detail at absolute zero for the simplest $d=1$ case, 
bearing the application to the
2D case in mind. 

In \S 2, we introduce the  model Hamiltonian. In \S 3, we study the
density of states based on the
perturbation theory with a special emphasis on both the forward 
and Umklapp scattering in the t-matrix approximation. 
Since we found that the t-matrix has
a pole in the upper half plane if the Umklapp term becomes dominant,  
we study the nature of this instability in the mean field 
approximation in \S 4. Another type of
instability associated with Umklapp
processes is studied here.
\S 5  is devoted to conclusion and discussion. 

In the present paper, we take unit of $\hbar=k_{B}=1$.
\section{Model Hamiltonian}
We start with the one-dimensional Hubbard model with 
nearest-neighbor hopping:\
\begin{equation}
H = -t\sum_{<ij>\sigma}\{c^{\dagger}_{i\sigma}c_{j\sigma}+(h.c.)\} +
U\sum_{i}n_{i\uparrow}n_{i\downarrow}-\mu\sum_{i}c^{\dagger}_{i\sigma}
c_{i\sigma},
\label{eqn:hubbard}
\end{equation}
where $t(>0)$ is the transfer integral, $c_{i\sigma}(c^{\dagger}_{i\sigma})$
is the annihilation (creation) operator for the electron on the i-th site
with spin $\sigma$, $U(>0)$ is the Coulomb repulsion
 and $\mu$ is the chemical potential.

\begin{figure}
\begin{center}
\leavevmode \epsfysize=5cm
\epsfbox{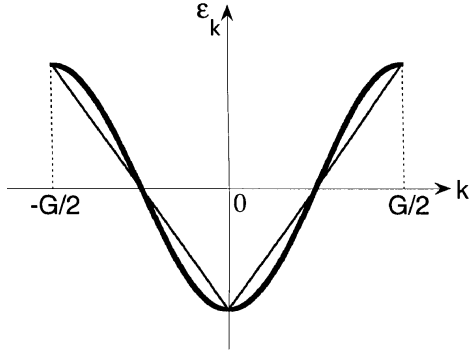}
\end{center}
\caption{Dispersion relation. Full and thin line show
the cosine band and the linearized one, respectively.}
\label{dispersion}
\end{figure}

We can classify normal and Umklapp processes by the Fourier transformation,
\begin{subequations}
\begin{equation}
H = H_{0} + H^{int} \label{eqn:hubbard2},
\end{equation}
\begin{equation}
H_{0} = \sum_{-G/2<k<G/2}\:\sum_{\sigma}\:\xi_{k}c^{\dagger}_{k\sigma}
c_{k\sigma},\label{eqn:hubbard3}
\end{equation}
\begin{equation}
\xi_{k}= \epsilon_{k}-\mu,
\end{equation}
\begin{equation} 
\epsilon_{k} = -2t\cos k,\label{eqn:cosine} 
\end{equation}
\begin{equation}
H^{int}=H^{int}_{N} + H^{int}_{U},
\end{equation}
\begin{full}
\begin{eqnarray}
\lefteqn{H^{int}_{N}}\nonumber\\ 
&=&\frac{U}{N}\sum_{0<q<G}\left\{\sum_{\stackrel{\scriptstyle{q-G/2<k<G/2}}
{q-G/2<k^{\prime}<G/2}}
c^{\dagger}_{k^{\prime}\uparrow}c^{\dagger}_{-k^{\prime}
+q\downarrow}c_{-k+q\downarrow}c_{k\uparrow}
+\sum_{\stackrel{\scriptstyle{-G/2<k<q-G/2}}
{-G/2<k^{\prime}<q-G/2}}
c^{\dagger}_{k^{\prime}\uparrow}c^{\dagger}_{-k^{\prime}
+q-G\downarrow}c_{-k+q-G\downarrow}c_{k\uparrow}\right\}, \label{eqn:hubbard4}\\
\lefteqn{H^{int}_{U}}\nonumber\\
&=& \frac{V}{N}\sum_{0<q<G}\left\{\sum_{
\stackrel{\scriptstyle{q-G/2<k<G/2}}{-G/2<k^{\prime}<q-G/2}}
c^{\dagger}_{k^{\prime}\uparrow}c^{\dagger}_{-k^{\prime}
+q-G\downarrow}c_{-k+q\downarrow}c_{k\uparrow}
+(h.c.)\right\}, \label{eqn:hubbard5}
\end{eqnarray}
\end{full}
\end{subequations}
where $N$, $R_{i}$, $G$ and $q$ (or $q-G$) are the number of lattice sites,  
the lattice site vector,
the reciprocal lattice vector 
equal to $2\pi$ and the total momentum of two electrons, respectively. 
Here the lattice constant is taken as unity. 
Each wave vector $k$ of the electron lies in the first Brillouin zone, 
$-G/2<k<G/2$. 
The cosine band 
is shown in Fig.~\ref{dispersion}. The bandwidth $W$ is equal to $4t$.
$H^{int}_{N}$ stands for normal processes which conserve the total
momentum $q$ of two electrons, and $H^{int}_{U}$ stands for
Umklapp processes in which it differs by $G$.
Since we are mainly interested in Umklapp processes, we set the magnitude of
interaction in $H^{int}_{U}$ as an arbitrary parameter $V(>0)$. 
In this way, 
we examine the effects of normal and Umklapp scattering 
separately.
\section{The Density of States in the Modified T-matrix Approximation}
In this section, we study the effects of the Umklapp scattering on the 
density of states based on the perturbation theory, at absolute zero $T=0$. 
For simplicity, we approximate the cosine dispersion with the liniarized one:
\begin{subeqnarray}
\epsilon_{k}=v(|k|-G/4),&&\\
\xi_{k}=v(|k|-k_{F}),\makebox[0.5em]{}\label{eqn:linear}
\end{subeqnarray}
where $k_{F}$ is the Fermi wave vector
and we put $v=8t/G$ so that the bandwidth $W$ is left unchanged. 
In the half-filled case, $k_{F}$ is equal to $G/4$. 
This dispersion is shown in Fig.~\ref{dispersion}.

\begin{figure}
\begin{center}
\leavevmode \epsfysize=2cm
\epsfbox{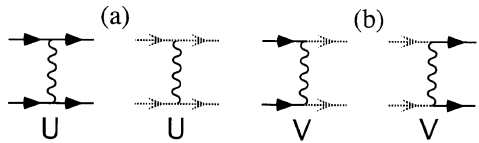}
\end{center}
\caption{(a) Forward and (b) Umklapp scattering processes. 
Full and dotted arrow lines stand for right-moving electron and 
left-moving electron, respectively.}
\label{fuscattering}
\end{figure}

We separate the dispersion into the branch with $k>0$ (right-moving)
and with $k<0$ (left-moving) 
and treat only the scattering
between two electrons on the same branch -
the forward and Umklapp scattering, as shown in 
Fig.~\ref{fuscattering}.
However, we neglect backward scattering processes in this paper, 
where the total momentum of two incident 
electrons is small.
\subsection{Particle-Particle Correlation Function and the T-Matrix}
\begin{figure}
\begin{center}
\leavevmode \epsfysize=2cm
\epsfbox{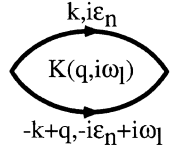}
\end{center}
\caption{Particle-particle correlation function, $K(q,i\omega_{l})$.}
\label{ppc}
\end{figure}

First of all we define the particle-particle correlation function 
$K(q,i\omega_{l})$
with the total momentum q and energy $i\omega_{l}$ shown in Fig.~\ref{ppc},
\begin{full}
\begin{subeqnarray}
K(q,i\omega_{l}) &=& T\sum_{i\epsilon_{n}}\sum_{k}G_{0}(k,i\epsilon_{n})
G_{0}(-k+q,-i\epsilon_{n}+i\omega_{l}), \\
&=& \sum_{k}\frac{f(\xi_{k})-f(-\xi_{-k+q})}
{i\omega_{l}-\xi_{k}-\xi_{-k+q}},\\
G_{0}(k,i\epsilon_{n}) &=& \frac{1}{i\epsilon_{n}-\xi_{k}},
\end{subeqnarray}
\end{full}
where $G_{0}$ is the thermal Green function of $H_{0}$, $\omega_{l}=2l\pi T$ 
and $\epsilon_{n}=(2n+1)\pi T$ are the Matsubara frequencies and 
$f(\xi)$ is the Fermi distribution function.
Since we consider only the interaction 
between two electrons on the same branch,
$\xi_{k}+\xi_{-k+q}$ is independent of k 
and we can calculate $K(q,i\omega_{l})$ analytically.
Especially, for $|q|\sim 2k_{F}$ at absolute zero $T=0$,
\begin{equation}
K(q,z)= -\frac{1}{2\pi v}\frac{x_{q}}{z-x_{q}},\label{eqn:kqz}\\
\end{equation}
where
\begin{equation}
x_{q} \equiv v(|q|-2k_{F}).
\end{equation}
We note that $K(q,z)$ has one pole on the real axis $z=x_{q}$ and
for $|q|\sim 2k_{F}$ the residue is small.

\begin{figure}
\begin{center}
\leavevmode \epsfysize=3.2cm
\epsfbox{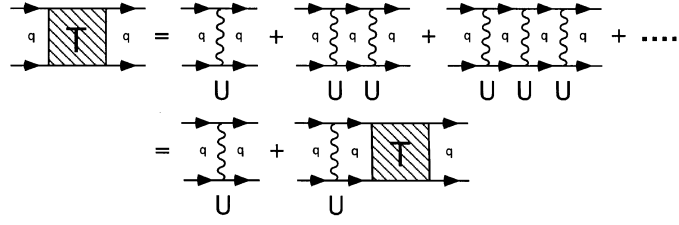}
\end{center}
\caption{The t-matrix, $T(q,i\omega_{l})$.}
\label{tmatrix}
\end{figure}

Next we define the t-matrix $T(q,i\omega_{l})$ without Umklapp processes 
(i.e., $V=0$),
\begin{equation}
T(q,i\omega_{l}) = -\frac{U}{1+UK(q,i\omega_{l})}.
\end{equation} 
This is the particle-particle ladder diagrams shown in Fig. ~\ref{tmatrix}. 
By use of $K(q,z)$, eq. (\ref{eqn:kqz}), we can calculate the t-matrix
analytically. Especially for $|q|\sim 2k_{F}$,
\begin{equation}
T(q,z) = -U-Ua\frac{x_{q}}{z-(1+a)x_{q}},\label{eqn:tmat}
\end{equation}
where
\begin{equation}
a\equiv \frac{U}{2\pi v}\equiv \frac{U}{2W}.
\end{equation}
We note that $T(q,z)$ also has one pole on the real axis
and for $|q|\sim 2k_{F}$ the residue is small.

\subsection{The Modified T-matrix}
\begin{figure}
\begin{center}
\leavevmode \epsfysize=6cm
\epsfbox{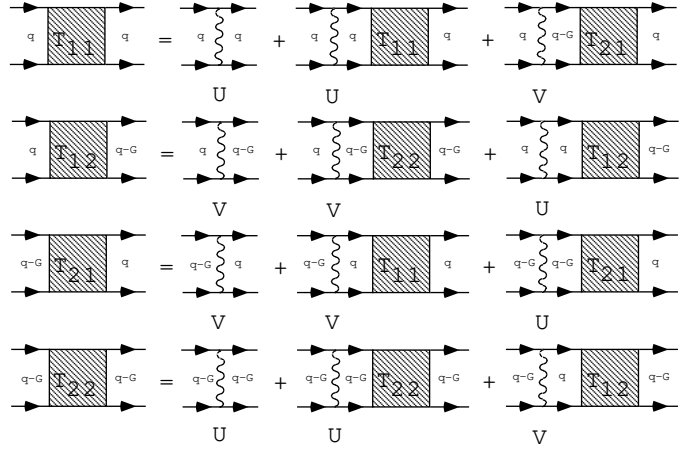}
\end{center}
\caption{The modified t-matrix, $\hat{T}(q,i\omega_{l})$.}
\label{mtmatrix}
\end{figure}

We consider the particle-particle ladder diagram including Umklapp processes
(i.e., $V\neq 0$). In this case the total momentum
q of incident two electrons can be changed into $q-G$ ($q>0$) 
after the Umklapp scattering, i.e., the Umklapp interaction introduces
the Bragg reflection of two incident electrons as a whole. 
In this case, the t-matrix is given as follows:
\begin{subequations}
\begin{equation}
\hat{T}\equiv -\hat{U}-\hat{U}\hat{K}\hat{T},
\end{equation}
\begin{equation}
\hat{T}= \left(
\begin{array}{cc}
T_{11}(q,z) & T_{12}(q,z)\\
T_{21}(q,z) & T_{22}(q,z)
\end{array}
\right),
\end{equation}
\begin{equation}
\hat{U}=\left(
\begin{array}{cc}
U & V\\
V & U
\end{array}
\right),
\end{equation}
\begin{equation}
\hat{K}= \left(
\begin{array}{cc}
K(q,z) & 0\\
0 & K(q-G,z)
\end{array}
\right),
\end{equation}
\end{subequations}
where the index 1 stands for the total momentum $q>0$ and 2 for $q-G<0$.
We call this matrix as a {\em modified} t-matrix.
This is the particle-particle ladder diagram shown in Fig.~\ref{mtmatrix}.
Especially for $q\sim 2k_{F}$, $T_{11}(q,z)$ is given by
\begin{full}
\begin{subequations}
\begin{equation}
T_{11}(q,z)= -U-\frac{U}{x_{+}(q)-x_{-}(q)}
\sum_{\gamma=\pm}\gamma\frac{(x_{\gamma}(q)-x_{q})(x_{\gamma}(q)-cx_{q-G})}
{z-x_{\gamma}(q)},
\end{equation}
\begin{equation}
x_{\pm}(q)\equiv\frac{1}{2}\left\{(1+a)(x_{q}+x_{q-G})\pm
\sqrt{(1+a)^{2}(x_{q}-x_{q-G})^{2}+4b^{2}x_{q}x_{q-G}}\right\},
\end{equation}
\begin{equation}
c \equiv (1+a)-\frac{b^{2}}{a},\makebox[1em]{}
b\equiv  \frac{V}{2\pi v}\equiv\frac{V}{2W},\makebox[4em]{}
\end{equation}
\end{subequations}
\end{full}
and has two poles $z=x_{\pm}(q)$. In the limit $V\rightarrow 0$,
 $T_{11}(q,z)$ reduces to eq. (\ref{eqn:tmat}).

In the half-filled case ($G=4k_{F}$), $x_{q-G}=-x_{q}$ and then 
\begin{equation}
x_{\pm}(q)=\pm\alpha x_{q},\makebox[1em]{}\alpha\equiv\sqrt{(1+a)^{2}-b^2},
\end{equation}
and therefore $T_{11}(q,z)$ for $q\sim 2k_{F}$ is written as
\begin{full}
\begin{eqnarray}
T_{11}(q,z) 
&=& -U-U\left \{\frac{(\alpha -1)(\alpha +c)}{2\alpha}\frac{x_{q}}{z-\alpha x_{q}}
-\frac{(\alpha +1)(\alpha -c)}{2\alpha}\frac{x_{q}}{z+\alpha x_{q}}\right \}.
\label{eqn:mtmat}
\end{eqnarray}
\end{full}
The renormalization factor $\alpha$ of the velocity $v$ becomes pure imaginary
if $b>1+a$, i.e., $V>U+2W$. Therefore a pole appears in the upper
half plane in this case. This fact implies that if
the Umklapp interaction is sufficiently strong, the system becomes unstable.
We will study this instability in the next section.
In this section, $\alpha$ is assumed to be real.

\subsection{The Density of States}
In this section, we calculate the density of states.
First of all, the single-particle spectral weight, $\rho (k,\epsilon)$, is given by
\begin{full}
\begin{subeqnarray}
\rho(k,\epsilon)&=&-\frac{1}{\pi}\mbox{Im}G^{R}(k,\epsilon),\\
&=& \rho_{c}(k,\epsilon) + \rho_{p}(k,\epsilon),\label{eqn:spweight}
\end{subeqnarray}
\end{full}
where $G$ is the full Green function with the proper self energy $\Sigma$,
$G^{R}(k,\epsilon)=[\epsilon-\xi_{k}-\Sigma^{R}(k,\epsilon)]^{-1}$.
The superscript R stands for the retarded function.
In eq. (\ref{eqn:spweight}),
$\rho_{c}(k,\epsilon)$ stands for the contribution from continuum part 
with $\mbox{Im}\Sigma^{R}(k,\epsilon)\neq 0$,
\begin{full}
\begin{equation}
\rho_{c}(k,\epsilon)=-\frac{1}{\pi}\frac{\mbox{Im}\Sigma^{R}
(k,\epsilon)}{\{\epsilon-\xi_{k}-\mbox{Re}\Sigma^{R}(k,\epsilon)\}^{2}
+\{\mbox{Im}\Sigma^{R}(k,\epsilon)\}^{2}},
\end{equation}
\end{full} 
and $\rho_{p}(k,\epsilon)$ for the contribution from the poles with
Im$\Sigma^{R}(k,\epsilon)=0$,
\begin{eqnarray}
\rho_{p}(k,\epsilon)&=&\delta(\epsilon-\xi_{k}-\mbox{Re}
\Sigma^{R}(k,\epsilon)),
\nonumber\\
&=&\sum_{i}z_{i}(\epsilon)\delta(k-k_{i}(\epsilon)), 
\end{eqnarray}
where
\begin{equation}
z_{i}(\epsilon)=\left | \frac{\partial}{\partial k}
[\xi_{k}+\mbox{Re}\Sigma^{R}(k,\epsilon)]\right |^{-1}_{k=k_{i}(\epsilon)},
\end{equation}
and $k_{i}(\epsilon)$ (i$=0,1,2,\cdots$) is the solution of
$\epsilon-\xi_{k}-\mbox{Re}\Sigma^{R}(k,\epsilon)=0$ satisfying 
Im$\Sigma^{R}(k,\epsilon)=0$
for given $\epsilon$. The summation is taken over all $k_{i}(\epsilon)$.

By use of $\rho(k,\epsilon)$, we obtain the density of states $\rho(\epsilon)$
per lattice site and spin,
\begin{equation}
\rho(\epsilon)=\rho_{c}(\epsilon)+\rho_{p}(\epsilon),
\end{equation}
where
\begin{equation}
\rho_{\nu}(\epsilon)=\frac{1}{N}\sum_{k}\rho_{\nu}(k,\epsilon),
\end{equation}	
and $\nu=c,p$.
In our approximation where the scattering processes 
between two electrons on the same branch are taken into account with the linear dispersion eq. (\ref{eqn:linear}),
we can calculate $\Sigma^{R}(k,\epsilon)$ 
analytically by use of the t-matrix in both cases of $V=0$ and $V\neq 0$. 
\subsubsection{The 2nd Order Perturbation in U}
\begin{figure}
\begin{center}
\leavevmode \epsfysize=5cm
\epsfbox{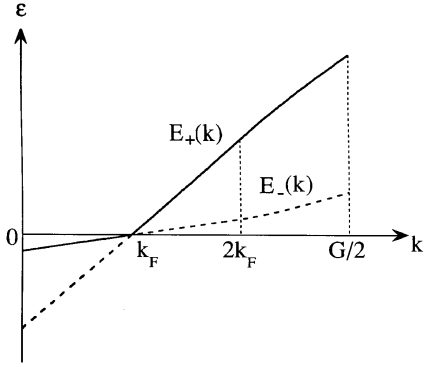}
\end{center}
\caption{The quasiparticle energy bands 
relative to the renormalized chemical potential
in the 2nd order perturbation in $U$ for $k>0$ and less-than-half-filled case. 
The states with $\epsilon <0$ are occupied.}
\label{2nddisp}
\end{figure}

First of all, we analyze the results in the 2nd order perturbation 
in $U$ but $V=0$. In this case,
the self energy can be written with $K(q,z)$ as follows:
\begin{equation}
\Sigma(k,i\epsilon_{n}) = -TU^{2}\sum_{i\omega_{l}}\sum_{q}
G_{0}(-k+q,-i\epsilon_{n}+i\omega_{l})K(q,i\omega_{l}).
\end{equation}
After straightforward calculations, we obtain for $k>0$,
\begin{equation}
\Sigma^{R}(k,\epsilon)=\left\{
\begin{array}{cc}
\frac{\displaystyle 1}{\displaystyle 2}a^{2}
\frac{\displaystyle\xi_{k}^{2}}{\displaystyle\epsilon-\xi_{k}+i\delta} &
0<k<2k_{F},\\
\frac{\displaystyle 1}{\displaystyle 2}a^{2}
\frac{\displaystyle\xi_{k}^{2}-\xi_{k-k_{F}}^{2}}
{\displaystyle\epsilon-\xi_{k}+i\delta} & 2k_{F}<k<G/2.
\end{array}
\right .
\end{equation}
Therefore it is seen that the Green function has two poles at $\epsilon=
E_{\pm}(k)$, 
\begin{equation}
E_{\pm}(k)=\left \{
\begin{array}{cl}
(1\mp\frac{\displaystyle a}{\displaystyle\sqrt{2}})\xi_{k} 
& \mbox{ for $0<k<k_{F}$,}\\ 
(1\pm\frac{\displaystyle a}{\displaystyle\sqrt{2}})\xi_{k} 
& \mbox{ for $k_{F}<k<2k_{F}$,}\\
\xi_{k}\pm a\sqrt{k_{F}(k-3k_{F}/2)} & \mbox { for $2k_{F}<k<G/2$.}
\end{array}
\right.\label{eqn:seconddisp}
\end{equation}
These energy bands for $k>0$ are shown in Fig.~\ref{2nddisp}. 
In the half-filled case, $2k_{F}=G/2$ and it is seen from eq. 
(\ref{eqn:seconddisp})
that these dispersions are linear in $k$
over the first Brillouin zone.

\begin{figure}
\begin{center}
\leavevmode \epsfysize=5cm
\epsfbox{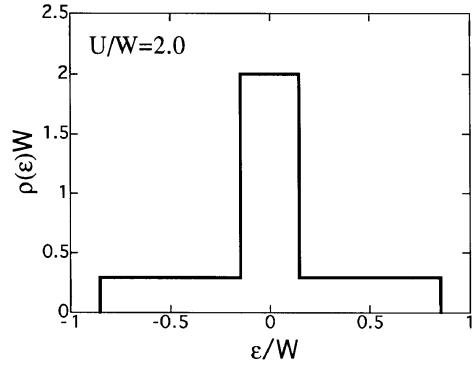}
\end{center}
\caption{The density of states in the second order
perturbation in $U$ for the case of half filling and $U/W=2.0$.}
\label{seconddos}
\end{figure}

The density of states consists only of
the contribution from the poles, $\rho_{p}(k,\epsilon)$,
which is shown in Fig.~\ref{seconddos} 
for the half-filled case with $U/W=2.0$. 
\subsubsection{The T-matrix Approximation for $U$}
\begin{figure}
\begin{center}
\leavevmode \epsfysize=5cm
\epsfbox{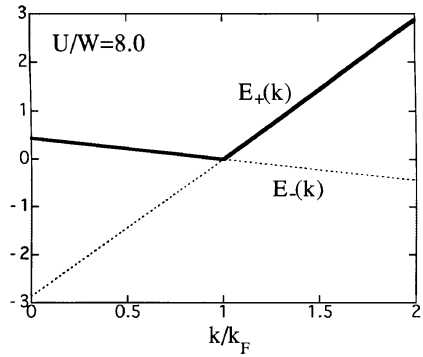}
\end{center}
\caption{The quasiparticle energy bands (scaled by $W$) relative to the 
renormalized chemical potential
in the t-matrix approximation for $0<k<2k_{F}$ and $U/W=8.0$.
The states with $\epsilon <0$ are occupied.}
\label{tmatdisp}
\end{figure}

Next we study the density of states in the 
t-matrix approximation for $U$ in the absence of $V$.
In this case, the self energy can be written with $T(q,z)$ as follows:
\begin{equation}
\Sigma(k,i\epsilon_{n}) = -T\sum_{i\omega_{l}}\sum_{q}
G_{0}(-k+q,-i\epsilon_{n}+i\omega_{l})T(q,i\omega_{l}).
\end{equation}
Especially for $0<k<2k_{F}$, $\Sigma^{R}(k,\epsilon)$ is given by
\begin{subequations}
\begin{equation}
\mbox{Re}\Sigma^{R}(k,\epsilon)=
-a\xi_{k}-(\epsilon-\xi_{k})\ln\left|\frac{\epsilon-(1+a)\xi_{k}}
{\epsilon-\xi_{k}}\right|,
\end{equation}
\begin{equation}
\mbox{Im}\Sigma^{R}(k,\epsilon)=
-\pi (\epsilon-\xi_{k})\mbox{sgn}(\xi_{k})\mbox{ for }
1<\epsilon / \xi_{k}<1+a.
\end{equation}
\end{subequations}
Here the contribution from the first term of the r.h.s. 
of (\ref{eqn:tmat}) is neglected, 
because this can be included in the chemical potential shift.
It is seen that the Green function $G^{R}(k,\epsilon)$ has two poles at
$\epsilon = E_{\pm}(k)$,
\begin{equation}
E_{\pm}(k)=\left \{
\begin{array}{cl}
c_{\mp}\xi_{k} & \mbox{ for $0<k<k_{F}$,}\\ 
c_{\pm}\xi_{k} & \mbox{ for $k_{F}<k<2k_{F}$,}
\end{array}
\right.
\end{equation}
where $c_{\pm}$ 
\begin{subeqnarray}
c_{+}=1+1.19a,\\
c_{-}=1-0.466a.
\end{subeqnarray}
are determined numerically.
There are two quasiparticle energy bands corresponding to these poles, which 
are shown for $0<k<2k_{F}$ in Fig.~\ref{tmatdisp}.

\begin{figure}
\begin{center}
\leavevmode \epsfysize=5cm
\epsfbox{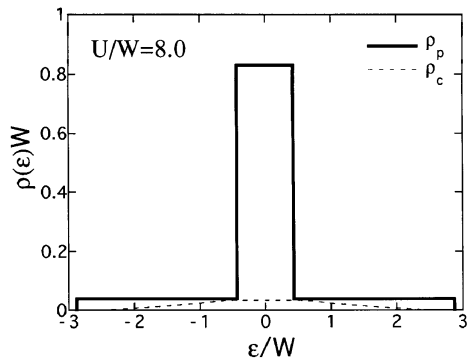}
\end{center}
\caption{The density of states in the t-matrix approximation for 
the case of half filling and $U/W=8.0$.}
\label{tmatdos}
\end{figure}

The resultant density of states, $\rho_{c}(\epsilon)$ and $\rho_{p}(\epsilon)$
in the half-filled case is shown 
in Fig.~\ref{tmatdos} for $U/W=8.0$.

In the t-matrix approximation, there exists the contribution to the spectral
weight also from the continuum part, i.e., $\rho_{c}(k,\epsilon)\neq 0$, 
although its contribution to the density of states, $\rho_{c}(\epsilon)$, 
is small.
It is seen that $\rho_{p}(\epsilon)$ has appreciable spectral weight
near the Fermi level $\epsilon\sim 0$. 
We note that for $U/W>>1$, $\rho(\epsilon)\neq 0$ over the range of 
$|\epsilon|$ \raisebox{-0.6ex}{$\stackrel{<}{\sim}$} $U/4$.

\subsubsection{The Modified T-matrix Approximation}
\begin{figure}
\begin{center}
\leavevmode \epsfysize=5cm
\epsfbox{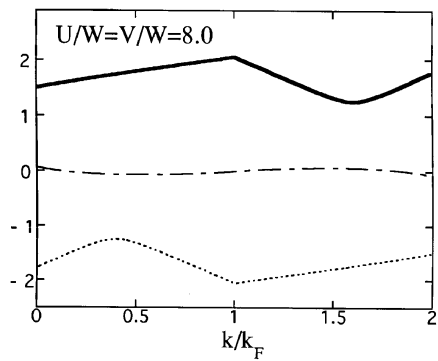}
\end{center}
\caption{The quasiparticle energy bands (scaled by $W$)
relative to the renormalized
chemical potential in the modified t-matrix approximation for the case of
half filling, $k>0$ and $U/W=V/W=8.0$. 
 The states with $\epsilon<0$ are occupied.}
\label{mtmatdisp}
\end{figure}

Finally, we include Umklapp processes by use of $T_{11}(q,z)$.
The modified t-matrix approximation 
leads to $\Sigma(k,i\epsilon_{n})$ for $k>0$ as given by
\begin{equation}
\Sigma(k,i\epsilon_{n}) = -T\sum_{i\omega_{l}}\sum_{q}
G_{0}(-k+q,-i\epsilon_{n}+i\omega_{l})T_{11}(q,i\omega_{l}).
\end{equation}
In this section, we will consider only the case of half filling 
for simplicity.
When we calculate $\Sigma(k,i\epsilon_{n})$ for $k>0$,
we take only the account of contributions of the $T_{11}(q,i\omega l)$
with $q\sim 2k_{F}$.
After straightforward calculations, we can evaluate 
$\Sigma^{R}(k,\epsilon)$ and then $\rho(k,\epsilon)$.
(The first term of the r.h.s. of (\ref{eqn:mtmat}) is included
in the chemical potential shift.)
There are three quasiparticle energy bands corresponding to the poles
of $G^{R}(k,\epsilon)$. These are shown in Fig.~\ref{mtmatdisp} 
for $k>0$ and $U/W=V/W=8.0$.
The Umklapp scattering produces the splitting between the upper and lower band,
between which there exists another energy band.

\begin{figure}
\begin{center}
\leavevmode \epsfysize=5cm
\epsfbox{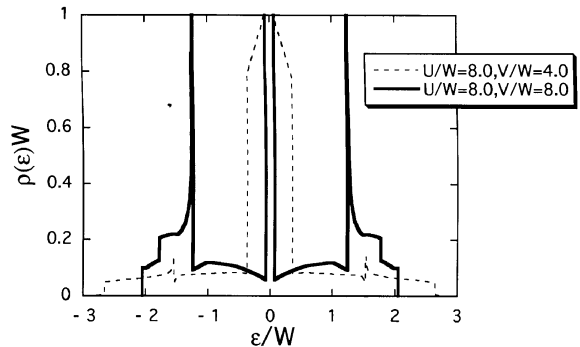}
\end{center}
\caption{The density of states in the half-filled case in the modified 
t-matrix approximation for $U/W=8.0$ and
$V/W=4.0$, $8.0$.} 
\label{mtmatdos1}
\end{figure}

The density of states is shown in Fig.~\ref{mtmatdos1}
for $U/W=8.0$ and a few values of $V/W$.
It is seen that for $V\neq0$, $\rho(\epsilon)$ has three-peak-structure. 
As the Umklapp scattering becomes strong, the upper and lower bands 
approach the Fermi level and the weight in these bands increases,
while the weight in the middle peak near the Fermi level decreases.

\begin{figure}
\begin{center}
\leavevmode \epsfysize=5cm
\epsfbox{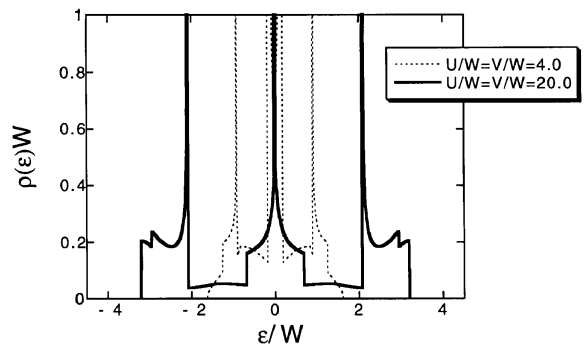}
\end{center}
\caption{The density of states in the modified t-matrix approximation for
the case of half filling and 
$U/W=V/W=4.0$, $20.0$.}
\label{mtmatdos2}
\end{figure}

The density of states in the case of $U=V$ is shown in 
Fig.~\ref{mtmatdos2}.
As $U$ or $V$ grows up in this case, the upper and lower band move away from
the Fermi level and three-peak-structure becomes prominent.
In this case, for $U/W>>1$, $\rho(\epsilon)\neq 0$ over the range of 
$|\epsilon|$ \raisebox{-0.6ex}{$\stackrel{<}{\sim}$} $\sqrt{U}/2$. 

As seen, Umklapp processes introduce the global splitting of the
density of states, although there remains the appriciable spectral weight
in the middle of split peaks and we cannot describe the Mott insulator.

We note that we calculated the self energy by the perturbation theory 
with the bare Green function
$G_{0}$. In this calculation, the system behaves like the Fermi liquid in that 
the renormalization factor $Z\neq0$.
We consider that this is due to the lack of the self-consistency.


But we emphasize that we are interested in the global feature of 
the density of states, $\rho(\epsilon)$, as a function of $\epsilon$
in the half-filled case
in the presence of the forward and Umklapp scattering.
\section{Mean Field Approximation}
In the last section, we found that if the Umklapp scattering becomes dominant, 
there appears a pole in the upper half plane in the modified t-matrix. 
In this section, we examine this instability 
in the mean field approximation.

We use the cosine band eq. (\ref{eqn:cosine}) as the dispersion. 
Hereafter, for simplicity, we set $U=0$ and consider $V$ only, 
and examine the ground state 
properties at absolute zero, $T=0$.  
\subsection{The $G/2-$pairing}
The fact that there exists an instability in the particle-particle
channel with the total momentum $|q|\sim 2k_{F}$ naturally
introduces the following order parameters,
\begin{subeqnarray}
\Delta_{R} \equiv \frac{V}{N}\mathop{{\sum}'}_{k}<c_{-k+G/2\downarrow}
c_{k\uparrow}>,\\
\Delta_{L} \equiv \frac{V}{N}\mathop{{\sum}'}_{k}<c_{-k\downarrow}
c_{k-G/2\uparrow}>,
\end{subeqnarray}
where 
\[
\mathop{{\sum}'}_{k} \equiv \sum_{0<k<G/2}.
\]
$\Delta_{R}(\Delta_{L})$ is the particle-particle pairing with the
total momentum $G/2(-G/2)$.
In the half-filled case, $G/2$ is equal to $2k_{F}$.\\
We extract $q=G/2$ term from $H^{int}_{U}$, which is decoupled to 
obtain the mean field Hamiltonian,
\begin{subequations}
\begin{equation}
H^{MF} = \sum_{\alpha =R,L}H_{\alpha}+H_{c},
\end{equation}
\begin{equation}
H_{\alpha} = \mathop{{\sum}'}_{k}
\psi^{\dagger}_{k\alpha}M_{k\alpha}\psi_{k\alpha},
\end{equation} 
\begin{equation}
H_{c}=-\frac{2}{V}\mbox{Re}(\Delta_{R}\Delta^{\ast}_{L})+\mathop{{\sum}'}_{k}
(\xi_{-k+G/2}+\xi_{-k}),
\end{equation}
\begin{equation}
\psi_{kR} = \left(
\begin{array}{c}
c_{k\uparrow}\\
c^{\dagger}_{-k+G/2\downarrow}
\end{array}
\right),
\psi_{kL} = \left(
\begin{array}{c}
c_{k-G/2\uparrow}\\c^{\dagger}_{-k\downarrow}
\end{array}
\right),
\end{equation}
\begin{equation}
M_{kR} = \left(
\begin{array}{cc}
\xi_{k} & \Delta_{L}\\
\Delta^{\ast}_{L} & -\xi_{-k+G/2}
\end{array}
\right),
M_{kL} = \left(
\begin{array}{cc}
\xi_{k-G/2} & \Delta_{R}\\
\Delta^{\ast}_{R} & -\xi_{-k}
\end{array}
\right).
\end{equation}
\end{subequations}
We note that there exists a self-consistent solution only for the case of
$\Delta_{R} = -\Delta_{L} \equiv \Delta_{1}$. 

\begin{figure}
\begin{center}
\leavevmode 
\epsfysize=4cm \epsfbox{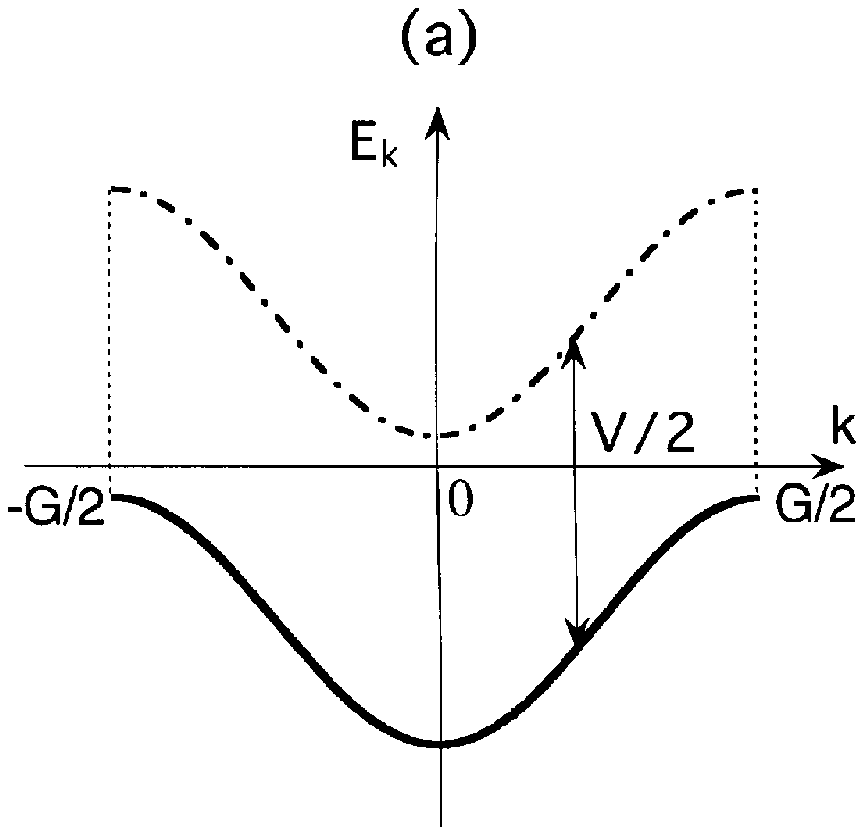}
\epsfysize=4.5cm \epsfbox{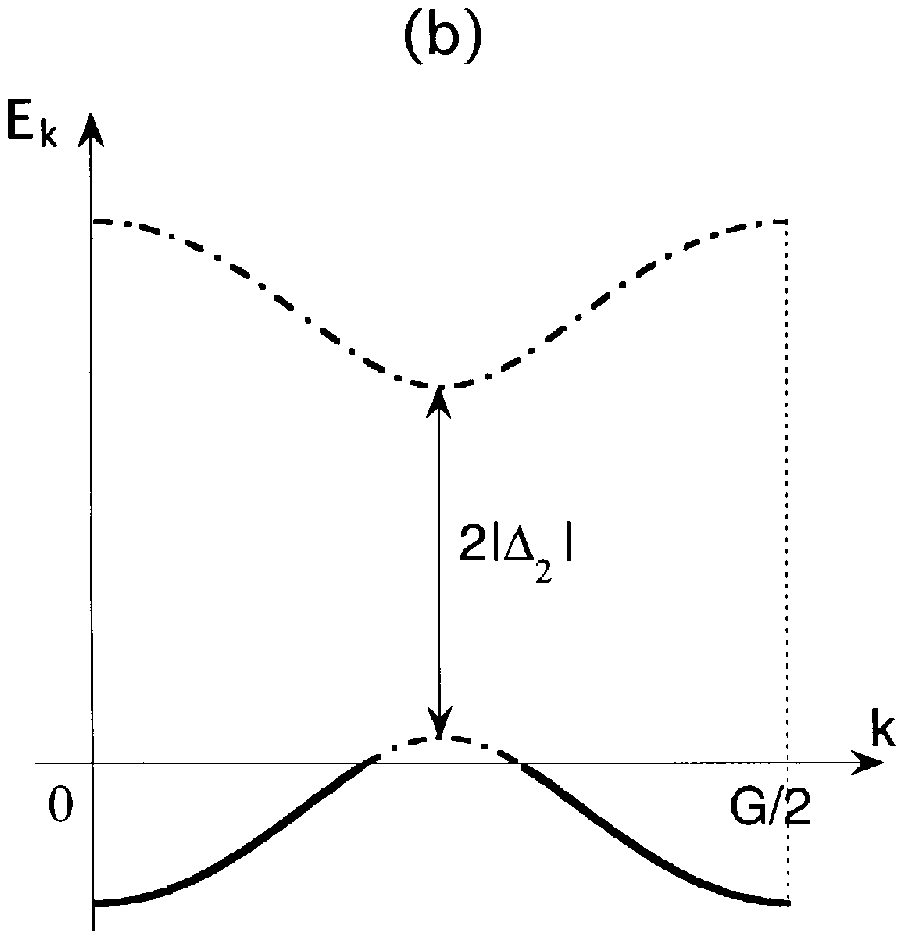}
\epsfysize=4.5cm \epsfbox{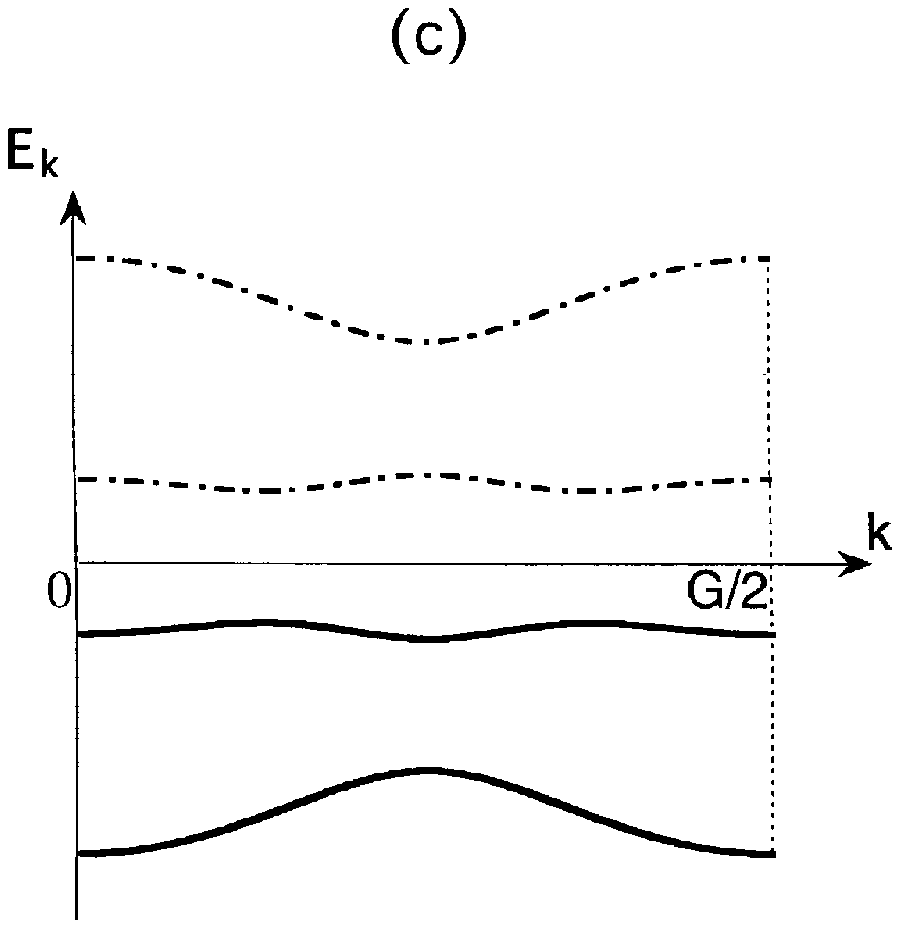}
\end{center}
\caption{The quasiparticle energy bands relative to the chemical potential
for (a)$G/2-$pairing state, (b)$G/2-$SDW state and 
(c) the coexisting state with both $G/2-$pairing and $G/2-$SDW.
Full line shows the occupied state.}
\label{qpdispersion}
\end{figure}

The quasiparticle dispersion in this case is given by
\begin{subeqnarray}
E^{\pm}_{\alpha}(k) = c_{\alpha}\epsilon_{k}\pm E_{g},\\
E_{g} \equiv \sqrt{\mu^{2}+|\Delta_{1}|^{2}}, 
\end{subeqnarray}
where $c_{\alpha}=1(-1)$ for $\alpha=R(L)$. 
This is shown in Fig.~\ref{qpdispersion}(a). 

It is seen that $\Delta_{1}\neq 0$ only for $V>8t$, in which case we have
\begin{eqnarray}
\mu&=&\frac{V}{4}(n-1),\label{eqn:mu1}\\
|\Delta_{1}|&=&\frac{V}{4}\sqrt{n(2-n)},\label{eqn:delta1}
\end{eqnarray}
where $n$ ($0<n<2$) is the filling (the expectation value of 
the number of electrons) per lattice site. 
(There is another solution for $8t<V<12\pi t$, although it corresponds to
a local maximum of the energy.) 
It turns out from eq. (\ref{eqn:mu1}) and (\ref{eqn:delta1}) that
$E_{g}\equiv V/4$ for $V>8t$ and the shape of the energy band does not
change for {\em arbitrary} filling $n$ 
and the Fermi level lies always in the energy gap. 
Such a state is possible, because the weight of the lower band 
shifts into the upper band as $n$ decreases.
If we use the linear dispersion, eq. (\ref{eqn:linear}), nonzero $\Delta_{1}$
exists for $V>2\pi v=2W$. Therefore it is seen that this 
corresponds to the condition that 
the modified t-matrix has a pole in the upper half plane.
The essential condition for the stability of this ordered state
is  $\epsilon_{k}+\epsilon_{-k+G/2}=0$.

The total energy per lattice site, $E_{1}$, for fixed $n$ is
\begin{equation}
E_{1}=\frac{V}{8}n(n-2).\label{eqn:totenergy1}
\end{equation}

We note that the existence of $\Delta_{1}$ does not lead to 
the $\eta$-pair condensation\cite{Eta1,Eta2}, since
\begin{eqnarray}
\hat{O}_{\eta -pair} &\equiv& \frac{1}{N}\sum_{i}(-1)^{i}c_{i\uparrow}c_{i\downarrow},\\
<\hat{O}_{\eta -pair}> &=& -\frac{1}{V}(\Delta_{R}+\Delta_{L}),\nonumber\\
&\equiv& 0.\nonumber
\end{eqnarray}
However the following type of coherence does exist,
\begin{eqnarray}
<c_{j\uparrow}c_{j\prime\downarrow}> &=& -i\frac{\sqrt{n(2-n)}}{\pi}
\frac{(-1)^{j}}{j-j\prime}\\
& &\mbox{for }j\prime=j\pm 1,j\pm 3,\cdots.\nonumber
\end{eqnarray}

\begin{figure}
\begin{center}
\leavevmode \epsfysize=5cm
\epsfbox{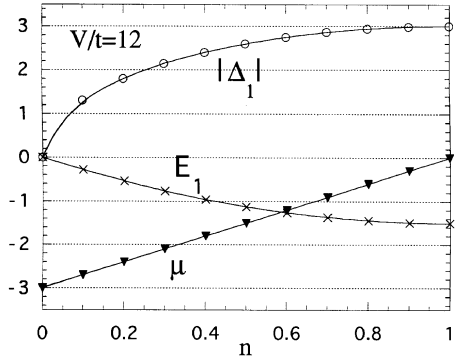}
\end{center}
\caption{The dependences of $|\Delta_{1}|$, chemical potential, $\mu$, 
and total energy, $E_{1}$, (scaled by $t$)
on the electron density in $G/2-$pairing state for $V/t=12$ .}
\label{delta1}
\end{figure}

The $n$-dependences of
$|\Delta_{1}|$, $\mu$ and $E_{1}$ are shown for $V/t=12$ in Fig.~\ref{delta1}.


\subsection{The $G/2-$SDW}
Since two electrons with $k\sim k_{F}$(or $-k_{F}$) are scattered 
into the final state with $k\sim -k_{F}$(or $k_{F}$),
the following particle-hole pairing order parameters are expected 
to be stabilized:
\begin{equation}
\Delta_{\sigma} \equiv \frac{V}{N}
\mathop{{\sum}'}_{k}<c^{\dagger}_{k-G/2\sigma}
c_{k\sigma}>,
\end{equation}
where $\sigma=\uparrow,\downarrow$ and $\mathop{{\sum}'}_{k} 
\equiv \sum_{0<k<G/2}$. 
The mean field Hamiltonian by using these order parameters
is given as follows,
\begin{subeqnarray}
H^{MF}= \sum_{\sigma}H_{\sigma}+H_{c},\makebox[6em]{}\label{eqn:mf2}&&\\
H_{\sigma} = \mathop{{\sum}'}_{k}
\psi^{\dagger}_{k\sigma}M_{k\sigma}\psi_{k\sigma}\mbox{, }
H_{c}=-\frac{2}{V}\mbox{Re}(\Delta_{\uparrow}\Delta_{\downarrow}),
\makebox[2em]{}&&\\
\psi_{k\sigma} = \left(
\begin{array}{c}
c_{k\sigma}\\
c_{k-G/2\sigma}
\end{array}
\right),
M_{k\sigma} = \left(
\begin{array}{cc}
\xi_{k} & \Delta^{\ast}_{\overline{\sigma}}\\
\Delta_{\overline{\sigma}} & \xi_{k-G/2}
\end{array}
\right),\mbox{}&&
\end{subeqnarray}
where $\overline{\sigma}$ stands for the spin opposite to $\sigma$.

Here we take
$\Delta_{\uparrow}=-\Delta^{\ast}_{\downarrow}\equiv \Delta_{2}$
so that there exists a self-consistent solution. We write $\Delta_{2}= 
|\Delta_{2}|e^{i\theta}$
where $0\leq \theta \leq \pi/2$. As we shall see later,
this phase factor is associated with the amplitude of the spin-density-wave
(SDW) order parameter.

The quasiparticle dispersion given by eq. (\ref{eqn:mf2})
is as follows,
\begin{subeqnarray}
E_{\pm}(k) = \pm E(k)-\mu,\\
E(k) \equiv \sqrt{\epsilon_{k}^{2}+|\Delta_{2}|^{2}}. 
\end{subeqnarray}
These are shown in Fig. ~\ref{qpdispersion}(b). In this case the 
energy gap is $2|\Delta_{2}|$.

We can determine $\Delta_{2}$ numerically from the self-consistent equation,
\begin{equation}
1=\frac{V}{2\pi}\int_{0}^{\frac{\pi}{2}(1-|1-n|)}\frac{dk}{E(k)},
\label{eqn:sce2}
\end{equation}
and the chemical potential $\mu$ is obtained by using this $\Delta_{2}$, 
\begin{equation}
\mu=\mbox{sgn}(n-1)\sqrt{|\Delta_{2}|^{2}+4t^{2}\cos^{2}(\frac{\pi}{2}n)}.
\end{equation}

This solution exists for $n_{c2}<n<2-n_{c2}$, where
\begin{equation}
n_{c2}=\frac{2}{\pi}\sin^{-1}[\tanh (\frac{4\pi t}{V})].\label{eqn:nc2}
\end{equation}
We see from eq. (\ref{eqn:sce2}) that in the half-filled case ($n=1$) 
this solution exists once $V>0$. In the limit ($V/t=\infty$), 
the energy gap is equal to $V/2$.
The nesting property $\epsilon_{k}+\epsilon_{k-G/2}=0$ is again crucial for the presence of this order parameter.

The total energy per lattice site, $E_{2}$, for fixed $n$ is
\begin{equation}
E_{2}=-\frac{2}{\pi}\int_{0}^{\frac{\pi}{2}(1-|1-n|)}E(k)dk
+2\frac{|\Delta_{2}|^{2}}{V}.\label{eqn:totenergy2}
\end{equation}

In accordance with nonzero $\Delta_{2}$, the long range order of SDW 
is induced, but not that of CDW, because
\begin{eqnarray}
\hat{O}_{CDW} &\equiv& \frac{1}{N}\sum_{i\sigma}(-1)^{i}c^{\dagger}_{i\sigma}
c_{i\sigma},\\
\hat{O}_{SDW} &\equiv& \frac{1}{2N}\sum_{i\sigma}(-1)^{i}
\sigma c^{\dagger}_{i\sigma}c_{i\sigma},\\
<\hat{O}_{CDW}> &=& \frac{2}{V}\mbox{Re}(\Delta_{\uparrow}+\Delta_{\downarrow}),\nonumber\\
&\equiv& 0,\\
<\hat{O}_{SDW}> &=& \frac{1}{V}\mbox{Re}(\Delta_{\uparrow}-\Delta_{\downarrow}),
\nonumber\\
&\equiv& 2\frac{|\Delta_{2}|}{V}\cos \theta.\label{eqn:SDW}
\end{eqnarray}
We note that the system is degenerate with respect to $\theta$ and 
in the case $\theta=\pi/2$ the SDW order vanishes.

Recently Wilczek {\em et.al.}
have introduced the same order parameter\cite{Wilczek}:
\[
<c^{\dagger}_{k+G/2\sigma}c_{k\sigma}> = \mbox{pure imaginary},
\]
which corresponds to $\theta=\pi/2$ in the present notation,
 i.e., without SDW order.
They showed that this order (which was named as 'd-Density order') 
exists in the 2-d extended Hubbard model with both on-site and
nearest-neighbor Coulomb repulsion, and have discussed the possible relationship
to the Mott insulators.
  
\begin{figure}
\begin{center}
\leavevmode \epsfysize=5cm
\epsfbox{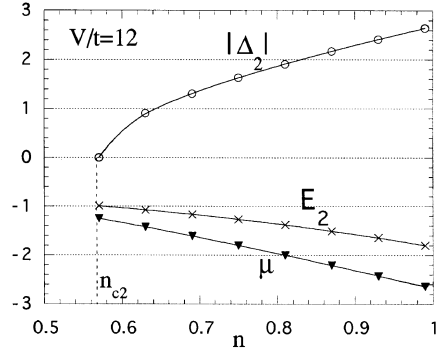}
\end{center}
\caption{The dependences of $|\Delta_{2}|$, chemical potential $\mu$ and total energy $E_{2}$ 
(scaled by $t$)
on the electron density in $G/2-$SDW state for $V/t=12$. 
For $n>n_{c2}$ $|\Delta_{2}|\neq 0$. In the half-filled case,  $\mu =0$.}
\label{delta2}
\end{figure}

The $n$-dependences of $|\Delta_{2}|$, $\mu$ and 
$E_{2}$ are shown for $V/t=12$ 
in Fig.~\ref{delta2}.
The important thing is that $E_{2}$ is decreasing function of $n$
, i.e., the compressibility $\kappa$ is negative, $\kappa\equiv\partial n/
\partial \mu <0$, except for the half-filled case $n=1$ at which  
$\mu$ changes discontinuously.
Therefore there occurs phase separation between a electron-rich and a 
electron-free phase.

\subsection{The $G/2-$pairing \&  $G/2-$SDW}
It is seen, in fact, that both $\Delta_{1}$ and $\Delta_{2}$ can
coexist.
To see this, we construct the mean-field Hamiltonian as follows:
\begin{subequations}
\begin{equation}
H^{MF}=\mathop{{\sum}'}_{k}
\psi^{\dagger}_{k}M_{k}\psi_{k}+H_{c},
\end{equation}
\begin{equation}
\psi^{\dagger}_{k}=(c^{\dagger}_{k\uparrow}\:c_{-k+G/2\downarrow}\:
c^{\dagger}_{k-G/2\uparrow}\:c_{-k\downarrow}),
\end{equation}
\begin{equation}
M_{k} = \left(
\begin{array}{cccc}
\epsilon_{k}-\mu & -\Delta_{1} & -\Delta_{2} & 0\\
-\Delta^{\ast}_{1} & \epsilon_{k}+\mu & 0 & -\Delta_{2}\\
-\Delta^{\ast}_{2} & 0 & -\epsilon_{k}-\mu & \Delta_{1}\\
0 & -\Delta^{\ast}_{2} & \Delta^{\ast}_{1}  & -\epsilon_{k}+\mu
\end{array}
\right),
\end{equation}
\begin{equation}
H_{c}=\frac{2}{V}(|\Delta_{1}|^{2}+|\Delta_{2}|^{2})-\mu.
\end{equation}
\end{subequations}
We note that, if both $\Delta_{1}$ and $\Delta_{2}$ are nonzero, 
the Cooper-pair also exists in general,
\begin{eqnarray}
\hat{O}_{C} &\equiv & \frac{1}{N}\sum_{i}c_{i\downarrow}c_{i\uparrow},\\
<\hat{O}_{C}>&=&\frac{1}{N}\sum_{-G/2<k<G/2}<c_{-k\downarrow}c_{k\uparrow}>
\label{eqn:Cooperpair}\\
&\neq & 0.\nonumber
\end{eqnarray}
Hence it is necessary, in principle, to consider the Cooper-pairing term 
in the mean field Hamiltonian.
In the present case of $U=0$, however,
such a term is
given by
\begin{eqnarray}
H&=&\frac{V}{N}\left \{ c^{\dagger}_{G/2\uparrow}c^{\dagger}_{G/2\downarrow}
\sum_{-G/2<k<G/2} c_{-k\downarrow}c_{k\uparrow}\:+\:(h.c.) \right \},
\nonumber \\
&\rightarrow& <c^{\dagger}_{G/2\uparrow}c^{\dagger}_{G/2\downarrow}>
\frac{V}{N}\sum_{-G/2<k<G/2}c_{k\downarrow}c_{-k\uparrow}\nonumber\\
& & \:+\:V<\hat{O}_{C}>
c^{\dagger}_{G/2\uparrow}c^{\dagger}_{G/2\downarrow}+(h.c.),\nonumber 
\end{eqnarray}
i.e., the term of the total momentum $q=0$ is associated only with
that of $q=G$ in Umklapp processes. Since the term $q=G$ consists of only one state 
of two electrons, each of which has wave vector $G/2$,
this gives the correction of $O(\frac{1}{N})$ to the total energy 
and we can neglect it.

In the presence of $\Delta_{1}$ and $\Delta_{2}$, 
the quasiparticle energies are given by $\pm E_{\pm}(k)$,
%
\begin{subeqnarray}
E_{\pm}(k)=\sqrt{\epsilon_{k}^{2}+\mu^{2}+|\Delta_{1}|^{2}+|\Delta_{2}|^{2}
\pm 2A(k)},\\
A(k)\equiv\sqrt{\epsilon_{k}^{2}|\Delta_{1}|^{2}+\mu^{2}(\epsilon_{k}^{2}+
|\Delta_{2}|^{2})}.\makebox[4em]{}
\end{subeqnarray}
This is shown in Fig.~\ref{qpdispersion}(c).

The self-consistent equations in this case are
\begin{full}
\begin{subeqnarray}
n&=&1+\mu\int_{0}^{\pi /2}\frac{dk}{\pi}\left\{ \frac{\epsilon_{k}^{2}+
|\Delta_{2}|^{2}}
{A(k)}[\frac{1}{E_{+}(k)}-\frac{1}{E_{-}(k)}]\:+\:
[\frac{1}{E_{+}(k)}+\frac{1}{E_{-}(k)}] \right\},\makebox[2em]{}
\label{eqn:sce31}\\
\Delta_{1}&=&V\Delta_{1}\int_{0}^{\pi /2}\frac{dk}{4\pi} \left\{
\frac{\epsilon_{k}^{2}}{A(k)}[\frac{1}{E_{+}(k)}-\frac{1}{E_{-}(k)}]\:+\:
[\frac{1}{E_{+}(k)}+\frac{1}{E_{-}(k)}] \right\},
\label{eqn:sce32}\\
\Delta_{2}&=&V\Delta_{2}\int_{0}^{\pi /2}\frac{dk}{4\pi} \left\{
\frac{\mu^{2}}{A(k)}[\frac{1}{E_{+}(k)}-\frac{1}{E_{-}(k)}]\:+\:
[\frac{1}{E_{+}(k)}+\frac{1}{E_{-}(k)}] \right\}\label{eqn:sce33}.
\end{subeqnarray}
\end{full}
These are solved numerically.
If there is a solution with $\Delta_{1},\Delta_{2}\neq0$, the correlation of
the Cooper-pair, eq. (\ref{eqn:Cooperpair}), is obtained as follows:
\begin{equation}
<\hat{O}_{C}>=-\frac{\mu}{V} 
|\Delta_{1}||\Delta_{2}| \sin \theta \int_{0}^{\pi /2}\frac{dk}{2\pi}
\frac{1}{A(k)}[\frac{1}{E_{+}(K)}-\frac{1}{E_{-}(k)}].\label{eqn:Cooper}
\end{equation}
We note that the phase factor, $\theta$, associated with 
the amplitude of SDW order parameter, eq. (\ref{eqn:SDW}),
appears also in the correlation of the Cooper-pair.

The solution with $\Delta_{1},\Delta_{2} \neq 0$ exists for $n_{c1}<n<2-n_{c1}$
 if $V>8t$,
where
\begin{equation}
n_{c1}\equiv 1-\sqrt{1-\sqrt{1-\left( \frac{8t}{V}\right) ^{2}}}.\label{eqn:nc1}
\end{equation} 
In the limit of $V\rightarrow 8t$, $n_{c1}\rightarrow 0$ and 
$\Delta_{1} \rightarrow 0$. It is seen that for a half-filled band $n=1$ ($\mu=0$),
this solution exists only in the limit $t=0$.   

\begin{figure}
\begin{center}
\leavevmode \epsfysize=5cm
\epsfbox{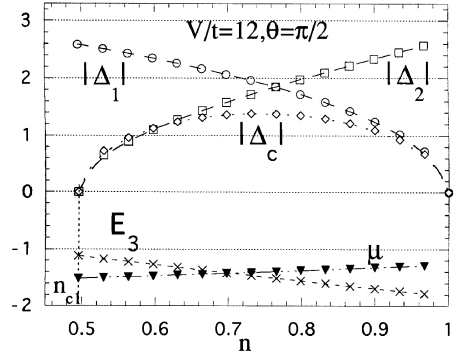}
\end{center}
\caption{The dependences of $|\Delta_{1}|$, $|\Delta_{2}|$, $|\Delta_{c}|$, chemical potential $\mu$ 
and total energy $E_{3}$ (scaled by $t$)
on the electron density in the coexisting
state with both $G/2-$pairing and SDW for $V/t=12$ and $\theta=\pi/2$ . These two states coexist for $n_{c1}<n<1$.}
\label{delta3}
\end{figure}

The total energy per lattice site, $E_{3}$, for fixed $n$ is
\begin{full}
\begin{equation}
E_{3}=-\int_{0}^{\pi /2}\frac{dk}{\pi} \left[ E_{+}(k)+E_{-}(k) \right]
+\frac{2}{V}(|\Delta_{1}|^{2}+|\Delta_{2}|^{2})+\mu (n-1).\label{eqn:totenergy3}
\end{equation}
\end{full}
The $n$-dependences of 
$|\Delta_{1}|$, $|\Delta_{2}|$, $|\Delta_{c}|\equiv V|<\hat{O}_{C}>|$,
$\mu$ and $E_{3}$ are shown in
Fig.~\ref{delta3} for $V/t=12$ and $\theta =\pi/2$. 
In the case of $\Delta_{1},\Delta_{2}
\neq 0$,
the compressibility $\kappa$ becomes positive in contrast to
the case of $\Delta_{1}=0$ and $\Delta_{2}\neq 0$. 

\subsection{Phase Diagram}
\begin{figure}
\begin{center}
\leavevmode \epsfysize=6cm
\epsfbox{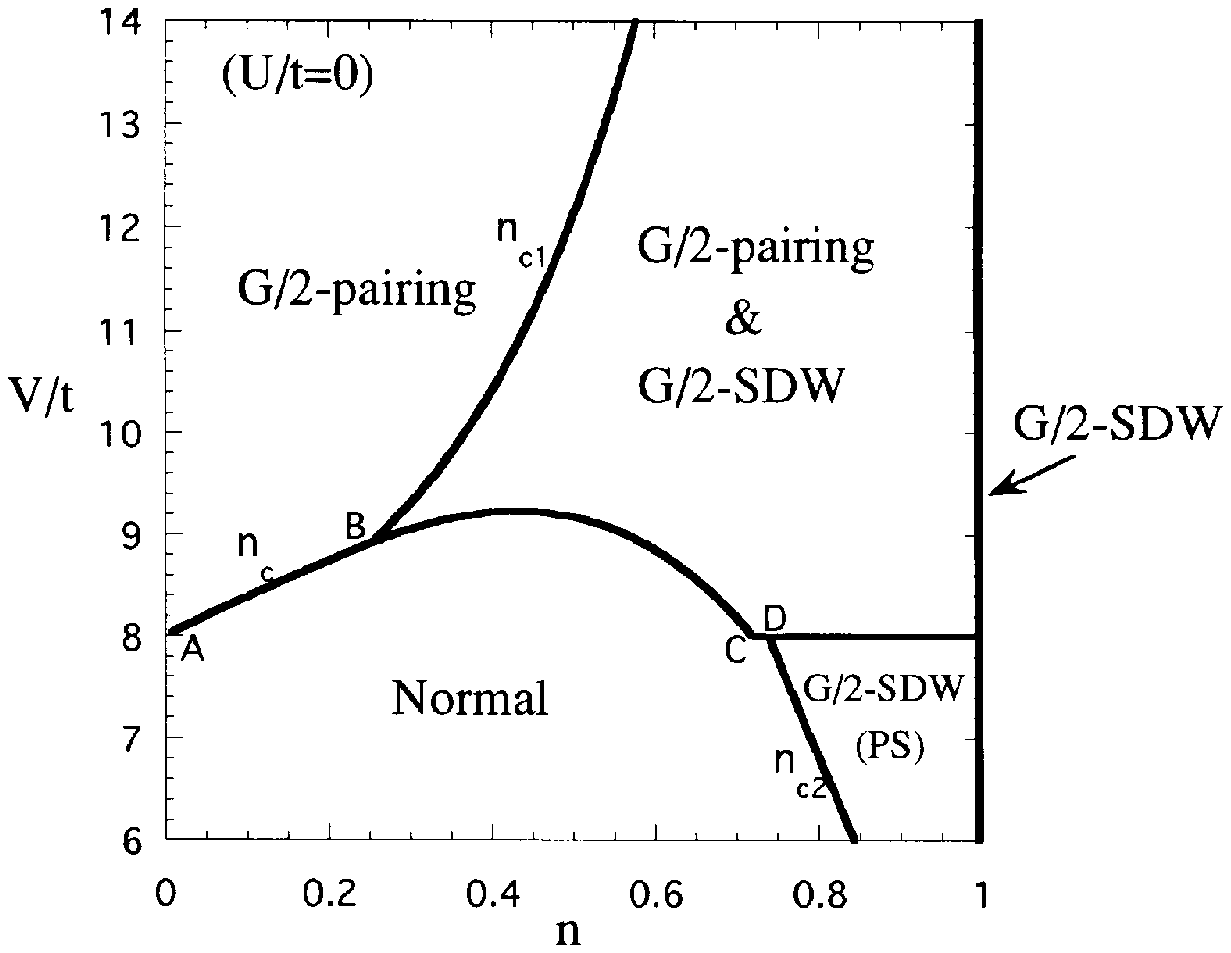}
\end{center}
\caption{Phase diagram in the plane of $V/t$ and $n$ for $U=0$.
PS stands for phase separation. Line AB stands for $n_{c}$.
Line BCD stands for the boundary between the normal state and
the coexisting state with both $G/2-$pairing and $G/2-$SDW.
For detailed explanations about $n_{c}$, $n_{c1}$ and $n_{c2}$,
see the text.}
\label{phasediagram}
\end{figure}

The state with the lowest energy has been determined in the $V/t-n$ plane
for the region of $0\leq n \leq 1$. The result is shown in 
Fig.~\ref{phasediagram}.

To begin with, we consider the case $V/t>8$. We compare $G/2-$pairing 
state with the normal state. 
By noting that the total energy of the normal state $E_{0}$ is
equal to $-\frac{4t}{\pi}\sin (\frac{\pi}{2}n)$ and comparing this with
$E_{1}$, eq. (\ref{eqn:totenergy1}), we see that the boundary between 
the normal state and the $G/2-$pairing state is of the first order 
as $V$ is varied and is given as follows,
\begin{equation}
\begin{array}{ll}
0<n<n_{c}&\mbox{for }8t<V<V_{c},\\
0<n\leq 1&\mbox{for }V>V_{c},
\end{array}
\end{equation}
where $V_{c}/t$ is equal to $32/\pi(\sim 10.2)$ and $n_{c}$ is determined
as the solution of the following equation,
\begin{equation}
\frac{V}{8}n_{c}(n_{c}-2)=-\frac{4t}{\pi}\sin (\frac{\pi}{2}n_{c}).
\end{equation}

Next we discuss $G/2-$SDW state. The transition from the normal state
to this state is of the second order as a function of $V$. 
As discussed in the last section,
this state is unstable against the phase separation (except 
for the case of half filling). However, it is energetically favorable compared
to the normal state, $E_{2}<E_{0}$ for $n_{c2}<n\leq 1$, where $E_{2}$ and $n_{c2}$
is given by eq. (\ref{eqn:totenergy2}) and eq. (\ref{eqn:nc2}), respectively.
On the other hand, this state competes 
with $G/2-$pairing state for $V>V_{c1}$ and $n_{c2}<n<n_{c1}$,  where
$V_{c1}/t\sim 13.0$ and $n_{c1}$ is given by eq. (\ref{eqn:nc1}).
($V_{c1}$ is determined as the solution of $n_{c1}=n_{c2}$.)
It is seen that in this region $G/2-$pairing state is stable, $E_{1}<E_{2}$.

In the coexisting state with both $G/2-$pairing and $G/2-$SDW for
$V/t>8$ and $n_{c1}<n<1$, 
the total energy, $E_{3}$, eq. (\ref{eqn:totenergy3}), is always lower than that of the state with only one order parameter among them, i.e.,
$E_{3}< E_{1},E_{2}$. On the other hand, 
for $V/t\sim 8$, the energy of this state is higher than
that of the normal state,
$E_{0}<E_{3}$, i.e.,
the transition from the normal state to this coexisting state is 
of the first order as a function of $V$. 
The boundary between the normal state and this coexisting state is determined 
numerically as the solution of $E_{0}=E_{3}$.
It is seen that $E_{3}<E_{0}$, independent of $n$, 
for $V/t$ \raisebox{-0.6ex}{$\stackrel{>}{\sim}$} $9.23$. 
Especially in this region, the ground state is 
given as follows:
\begin{equation}
\begin{array}{cc}
n=1&G/2-\mbox{SDW},\\
n_{c1}<n<1&G/2-\mbox{pairing \& }G/2-\mbox{SDW},\\
0<n<n_{c1}&G/2-\mbox{pairing},
\end{array}
\end{equation}

On the other hand, for $0<V/t<8$, the ground state is given as follows:
\begin{equation}
\begin{array}{cc}
n=1&G/2-\mbox{SDW},\\
0<n<n_{c2}&\mbox{normal},
\end{array}
\end{equation}
and for $n_{c2}<n<1$ there occurs phase separation.
\section{Conclusion and Discussion}
We have studied the effects of the Umklapp scattering  
in the 1D Hubbard model by treating the strength of normal ($U$) and Umklapp ($V$) 
interaction as independent parameters, 
keeping the bandwidth ($W$) finite. 

We calculated the density of states 
by use of the perturbation theory with the linear dispersion
with a special emphasis on both the forward and Umklapp scattering.
 
For $V=0$, we found that the t-matrix (the summation of
the particle-particle ladder diagram with respect to $U$) 
has one pole on the real axis. 
The density of states, $\rho(\epsilon)$, in the t-matrix approximation, 
has appreciable spectral weight and no energy gap at the Fermi level 
in the half-filled case. 

In the presence of $V$, we found that the {\em modified} t-matrix  
(including $V$ term) has two poles on the real axis for $0<V\leq U$. 
In the resultant $\rho(\epsilon)$ in the modified t-matrix approximation,
the global splitting is introduced and three-peak-structure is observed
in the half-filled case, although the appreciable spectral weight 
remains in the middle of split peaks near the Fermi level.
As the Umklapp scattering becomes strong, this structure becomes prominent.
However, we have not described the Mott insulator.

On the other hand, if $V>U+2W$, a pole of 
the modified t-matrix appears in the upper half plane.
Therefore in \S 4, first of all we studied this instability 
in the mean field approximation for $U=0$ and
found that in the strong coupling regime, $G/2-$pairing state becomes stable.
Next we examined another instability associated with Umklapp processes and
found that there appears $G/2-$SDW state. 
This state, however, 
is metastable due to the negative compressibility 
except for a half-filled band.
Finally, we found that the above two order parameters can coexist. 

We constructed a phase diagram in the full $V/t-n$ plane 
for these possible states.
In the strong coupling regime $V/t$  \raisebox{-0.6ex}{$\stackrel{>}{\sim}$} 
$8$, the ground state is the $G/2-$SDW state
at just half filling $n=1$, the coexisting state with both $G/2-$pairing 
and $G/2-$SDW near half filling and the $G/2-$pairing state at low
filling.
In this coexisting state near half filling, the correlation of 
the Cooper-pair appears 
and is connected with the SDW order parameter through one phase angle
, $\theta$, i.e., $<\hat{O}_{C}>\propto \sin \theta$, eq. (\ref{eqn:Cooper}),
and $<\hat{O}_{SDW}>\propto \cos \theta$, eq. (\ref{eqn:SDW}),
through the existence of the $G/2-$pairing.
Such close relationship between the SDW and Cooper-pair  
has been indicated in the 2D $t-J$ model with the 
SO(5) symmetry\cite{SCZhang},
which unifies antiferromagnetism with d-wave superconductivity.
The application of the present theory to such 2D case is 
the future problem.
\section*{Acknowledgements}
M.M. would like to express his gratitude to Hiroshi Kohno for instructive 
discussions and suggestions.  This work is financially supported by a Grant-in-Aid
for Scientific Research on Priority Area "Anomalous Metallic State near the 
Mott Transition" (07237102) from the Ministry of Education, Science, Sports
and Culture.

\end{document}